\documentclass[aps,prl,twocolumn,superscriptaddress]{revtex4}

\usepackage{graphicx}
\usepackage{amsmath}
\usepackage{amssymb}

\begin{document}

\title{Vortex Lattices in the Bose-Fermi Superfluid Mixture}
\author{Yuzhu Jiang}
\affiliation{State Key Laboratory of Magnetic Resonance and Atomic and Molecular Physics,
Wuhan Institute of Physics and Mathematics, Chinese Academy of Sciences, Wuhan 430071, China}
\author{Ran Qi}
\email{qiran@ruc.edu.cn}
\affiliation{Department of Physics, Renmin University of China, Beijing, 100872, China}
\author{Zhe-Yu Shi}
\affiliation{Institute for Advanced Study, Tsinghua University, Beijing, 100084, China}
\author{Hui Zhai}
\email{hzhai@tsinghua.edu.cn}
\affiliation{Institute for Advanced Study, Tsinghua University, Beijing, 100084, China}

\date{\today }

\begin{abstract}

In this letter we show that the vortex lattice structure in the Bose-Fermi superfluid mixture can undergo a sequence of structure transitions when the Fermi superfluid is tuned from the BCS regime to the BEC regime. This is due to different vortex core structure of the Fermi superfluid in the BCS regime and in the BEC regime. In the former the vortex core is nearly filled, while the density at the vortex core gradually decreases until it empties out at the BEC regime. Therefore, with the density-density interaction between the Bose and the Fermi superfluids, the two sets of vortex lattices interact stronger in the BEC regime that yields the structure transition of vortex lattices. In view of recent realization of this superfluid mixture and vortices therein, our theoretical predication can be verified experimentally in near future. 

\end{abstract}

\maketitle

Vortices play a fundamental and important role in superconductors and superfluids. In a type-II superconductor, as first discussed by Abrikosov, magnetic field can penetrate into the superconductor and form a triangular lattice \cite{Abrikosov}. Stable quantized vortices are also hallmark of the superfluidity of cold atomic gases, which can be created by rotation \cite{KetterleBEC, Dalibard,review} or synthetic magnetic field \cite{synthetic}, and triangular vortex lattices have been observed in both atomic Bose condensate \cite{KetterleBEC, Dalibard,review} and the Fermi superfluid across the BEC-BCS crossover \cite{KetterleFermion}. Vortex lattice has also been studied in the two-component atomic BEC \cite{HoMueller,Ueda}. It has been predicted that, due to the repulsion between atoms of different components, vortex cores in one component try to avoid overlapping with these in the other component, which leads to a sequence of structure transitions of vortex lattices as the repulsion increases, and eventually yields two sets of staggered rectangular lattices \cite{HoMueller,Ueda}. This phenomenon has been later observed experimentally \cite{JILA}.

Another important latest development in the superfluidity study is the realization of a Bose-Fermi superfluid mixture \cite{Salamon, Salamon2, Yuao}. In this system, bosons are weakly interacting and they are Bose condensed. Spin-$1/2$ fermions are in the strongly interacting regime and the scattering length between fermions is magnetic field tunable such that the Fermi superfluid can be controlled from the BCS regime to the BEC regime. In addition, there is also a repulsion between bosons and both two spin components of fermions. This interesting system has drawn lots of theoretically attention recently \cite{Zheng,Zhang,Cui,Stringari}. In the BEC limit the system can be viewed as a mixture of molecule BEC and atomic BEC, and therefore, the situation is similar to the two-component BEC studied before. However, the situation can be very different in the BCS side, it is the pairing order parameter that vanishes at the vortex core while the fermion density remains finite there. This is strongly in contrast to the BEC limit where both the order parameter and the density vanish simultaneously  at the vortex core. Since the interaction between bosons and fermions is density-density interaction, this difference will manifest itself significantly in determining the vortex lattice structure. The goal of this letter is to study the vortex lattice structure of the Bose-Fermi superfluid mixture for various parameters, in particular, when the interaction between fermions varies across the BEC-BCS crossover. Very recently, vortices have been successfully generated in the Bose-Fermi superfluid mixture \cite{Yuao}, therefore, our predication can be verified by the ongoing experiments in near future. 

\textit{Theory Framework.} The basic framework of our theory works as follows:

i) We consider a uniform system with fixed chemical potentials. That means we consider a two-dimensional isotropic trap whose frequency $\omega$ equals to the rotational frequency $\Omega$. In practice, $\omega$ is larger than $\Omega$ that gives rise to a residual trapping potential. This residual potential, together with the trapping potential along the third $z$-direction, can be treated by the local density approximation and we will not discuss this explicitly here in this work. For a uniform system with bulk density of bosons or fermions (denoted by $n_\text{b}$ or $n_\text{f}$), there is a unique relation between the chemical potential of bosons or fermions ($\mu_\text{b}$ or $\mu_\text{f}$) and the density. At the mean-field level, for bosons, $\mu_\text{b}=g_\text{bb}n_\text{b}$; and for fermions, the relation between $\mu_\text{f}$ and $n_\text{f}$ is given by the BEC-BCS crossover mean-field theory, as we will show below. Here $g_\text{bb}=4\pi \hbar^2 a_\text{bb}/m_\text{b}$ is the interaction strength between bosons ($m_\text{b}$ is the mass of bosons and $a_\text{bb}$ is the s-wave scattering length between bosons) . 

ii) We take the lowest Landau level approximation. With this approximation, we can use the Jacobi theta-function as the variational wave function for describing vortex lattices in the order parameters of both superfluids (that is, condensate wave function $\varphi$ for the Bose superfluid and the pairing order parameter $\Delta$ for the Fermi superfluid) \cite{Abrikosov,HoMueller}. The advantage of using the Jacobi theta-function is that, as we will show explicitly later, when the parameters of the vortex density and the unit vectors of the lattice are given, the entire function forms of $\varphi$ and $\Delta$ are determined analytically. Here we assume that the vortex densities in the Bose and the Fermi superfluids are two independent parameters. This is practically reasonable because when an external potential, say, a repulsive potential created by a laser, is applied to the system, the forces experienced by different atoms are generically different, and therefore the angular momentum deposited into two superfluids are different. 

iii) We determine the vortex lattice structure by minimizing the total free energy. Here the major assumption is that the local free energy density as a function of $\Delta({\bf r})$ and $\varphi({\bf r})$ takes the same form as the bulk system. This assumption works when the distance between vortices, or equivalent to say, the scale at which the order parameters vary, is much larger than the inter-particle distance. Hence, the total free energy contains three parts, $\mathcal{F}=\mathcal{F}_\text{b}+\mathcal{F}_\text{f}+\mathcal{F}_\text{bf}$. $\mathcal{F}_\text{b}$ is the free energy for bosons alone, which is given by 
\begin{equation}
\mathcal{F}_\text{b}=\int d^3{\bf r}\left( \frac{2\pi \hbar^2 a_\text{bb}}{m_\text{b}}n^2_\text{b}({\bf r})-\mu_\text{b}n_\text{b}({\bf r})\right),
\end{equation}
where $n_\text{b}({\bf r})=|\varphi({\bf r})|^2$ is the density of bosons, the free energy of fermion $\mathcal{F}_\text{f}$ is a functional of $\Delta({\bf r})$, and will be discussed in detail below. Finally, $\mathcal{F}_\text{bf}$ describes the density-density interaction between bosonic and fermionic atoms as 
\begin{equation}
\mathcal{F}_\text{bf}=\frac{2\pi\hbar^2 a_\text{bf}}{m_\text{bf}}\int d^3{\bf r}n_\text{b}({\bf r})n_\text{f}({\bf r}),
\end{equation}
where the s-wave scattering length between bosons and two different spin component of fermions is taken to be the same, as in the case of experiment \cite{Salamon,Yuao}, and denoted by $a_\text{bf}$, and $m_\text{bf}$ is the reduced mass between a boson and a fermion atom. Here the key is to determine the local density distribution of the Fermi superfluid $n_\text{f}({\bf r})$ as $\Delta({\bf r})$ varies. Once all these relations are known, the total free energy can be uniquely determined by $\varphi({\bf r})$ and $\Delta({\bf r})$; and as discussed in ii), since now $\varphi({\bf r})$ and $\Delta({\bf r})$ are uniquely determined by the vortex lattice vectors, hence, by minimizing the free energy, one can determine the entire lattice structure for both superfluids.

\begin{figure}[tp]
\includegraphics[width=3.2 in]
{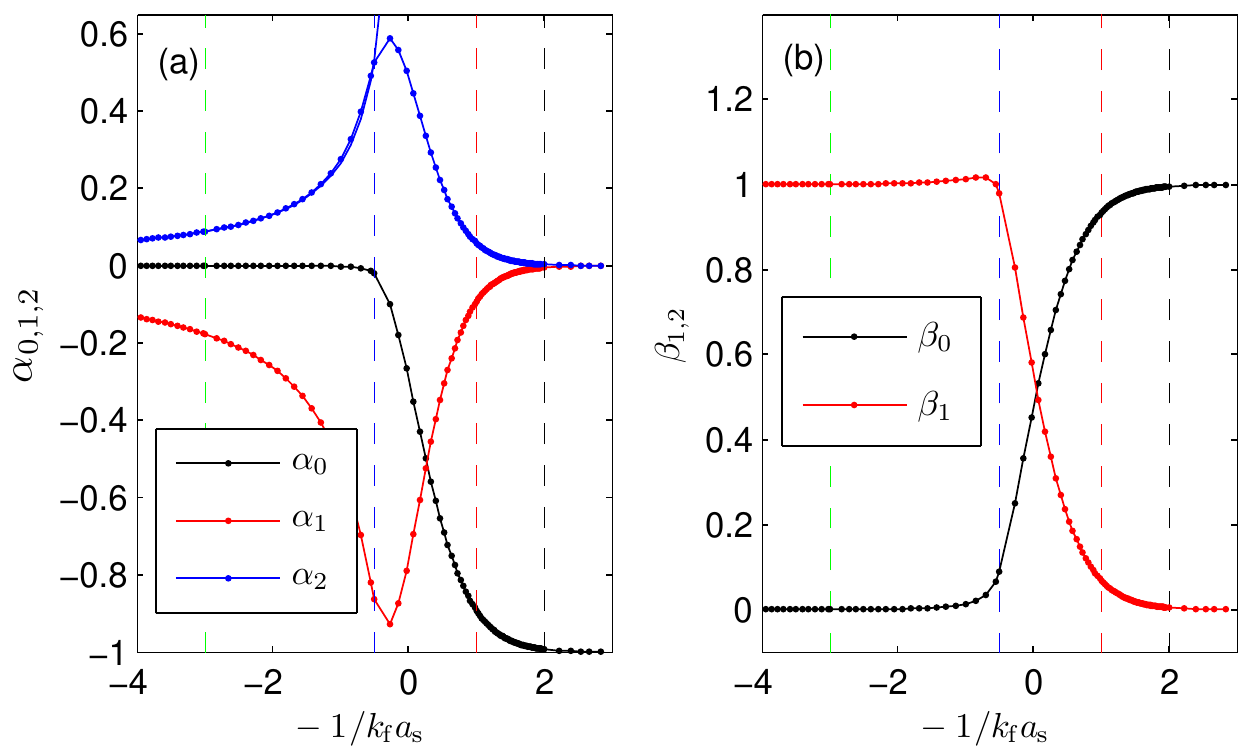}
\caption{The expansion coefficients $\alpha_{0,1,2}$ in Eq. \ref{Ff} (a) and $\beta_{0,1}$ in Eq. \ref{nf} as functions of $-1/(k_\text{F}a_{\text{s}})$   }
\label{alpha_beta}
\end{figure}

\textit{Some Required Relations for the Fermi Superfluid.} As discussed above, in order to calculate the free energy when $\Delta({\bf r})$ varies spatially, we need to know (a) $\mathcal{F}_\text{f}$ as a function of $\Delta$; and (b) $n_\text{f}$ as a function of $\Delta$. And for the convenience of later calculations, we need these relations in an expansion form as
\begin{align}
&\mathcal{F}_\text{f}=\mathcal{E}_\text{F}(\alpha_0+\alpha_1\bar{\Delta}^2+\alpha_2\bar{\Delta}^4+\dots), \label{Ff} \\
&n_\text{f}=n_{\text{f},0}(\beta_0+\beta_1\bar{\Delta}^2+\dots), \label{nf}
\end{align}
where $n_{\text{f},0}$ denotes the bulk density of fermions, $k_\text{F}=(6\pi^2 n_{\text{f},0})^{1/3}$ is the Fermi momentum, $\mathcal{E}_\text{F}=\hbar^2 k^5_\text{F}/(15m_\text{f}\pi^2)$ and $E_\text{F}=\hbar^2 k^2_\text{F}/(2m_\text{f})$ are the energy density for the free Fermi gas and the Fermi energy, respectively, and will be taken as energy units hereafter, where $m_\text{f}$ is the mass of fermion atoms.  $\bar{\Delta}=\Delta/\Delta_0$, and $\Delta_0$ is the bulk gap value. Here what we need to do is to determine $\alpha_{0,1,2}$ and $\beta_{0,1}$ as a function $-1/(k_\text{F}a_\text{s})$, and we determine them by fitting the numerical results from the BEC-BCS crossover mean-field theory, as shown in Fig. \ref{alpha_beta}. 

The mean-field theory for the BEC-BCS crossover contains the gap and the number equations as follows:
\begin{align}
&\frac{m_\text{f}}{4\pi\hbar^2 a_\text{s}}=\sum\limits_{{\bf k}}\left(\frac{1}{2\sqrt{(\epsilon_{{\bf k}}-\mu)^2+\Delta^2}}-\frac{1}{2\epsilon_{{\bf k}}}\right) \label{gap}\\
&n_{\text{f},0}=\sum\limits_{{\bf k}}\frac{1}{2}\left(1-\frac{\epsilon_{{\bf k}}-\mu}{\sqrt{(\epsilon_{{\bf k}}-\mu)^2+\Delta^2}}\right) \label{number},
\end{align}
where $\epsilon_{{\bf k}}=\hbar^2 {\bf k}^2/(2m_\text{f})$ and $a_\text{s}$ is the s-wave scattering length between fermions. 
From Eq. \ref{gap} and \ref{number}, we can determine the value of bulk gap $\Delta_0/E_\text{F}$ and $\mu/E_\text{F}$ as functions of $-1/(k_\text{F}a_\text{s})$, which can be found in many references \cite{mean-field} and therefore we will not display them here. Fixing $\mu$ and varying $\Delta$ in the range $0<\Delta<\Delta_0$, the L.H.S. of Eq. \ref{number} gives the fermion density $n_\text{f}/n_{\text{f},0}$ as a function of $\Delta/\Delta_0$ for any given $-1/(k_\text{F}a_\text{s})$. By fitting this function, one obtains the expansion Eq. \ref{nf} and the coefficients $\beta_{0,1}$ as functions of $-1/(k_\text{F}a_\text{s})$, which is shown in Fig. \ref{alpha_beta}(b). As one can see, in the BCS regime, $\beta_0\approx 1$ and $\beta_1\approx 0$, which means that the density remains a constant even when the order parameter $\Delta$ vanishes at the vortex core. Toward the BEC side, $\beta_0$ decreases and $\beta_1$ increases. Eventually, at the BEC regime, $\beta_0\approx 0$ and $\beta_1\approx 1$, which means that the density is proportional to the order parameter and vanishes simultaneously with the order parameter at the vortex core. We note that this difference plays an important role in the later conclusion of vortex lattice structure transition across the BEC-BCS crossover.

The free-energy is given by
\begin{equation}
\mathcal{F}_\text{f}=\sum_{{\bf k}}\left[(\epsilon_{{\bf k}}-\mu)-\sqrt{(\epsilon_{{\bf k}}-\mu)^2+\Delta^2}\right]+\frac{\Delta^2}{g_\text{f}} \label{free-energy}
\end{equation}
where 
\begin{equation}
\frac{1}{g_\text{f}}=-\frac{m_\text{f}}{4\pi\hbar^2 a_\text{s}}+\sum_{{\bf k}}\frac{1}{2\epsilon_{{\bf k}}}.
\end{equation}
For a given $\mu$, $\Delta_0$ minimizes this free-energy. While if one varies $\Delta$, the L.H.S. of Eq. \ref{free-energy} gives $\mathcal{F}_\text{f}/\mathcal{E}_\text{F}$ as a function of $\Delta/\Delta_0$ for each given $-1/(k_\text{F}a_\text{s})$. Again, by fitting this function, one obtains the expansion Eq. \ref{Ff} and the coefficients $\alpha_{0,1,2}$ as functions of $-1/(k_\text{F}a_\text{s})$ are shown in Fig. \ref{alpha_beta}(a). Here $\alpha_2$ represents the interaction between pairs, which is peaked at the unitary regime and decreases toward the BEC side.

\textit{Jacobi-Theta-Function as Variational Wave Functions.} Vortex lattice is described by a two-dimensional complex wave function whose zeros form a lattice (here we only consider simple Bravais lattice), and we denote the two lattice vectors as ${\bf b}_1= a \hat{x}$ and ${\bf b}_2=a(u\hat{x}+v\hat{y})$. $v_\text{c}=a^2 v$ denotes the area of a unit cell and $1/v_\text{c}$ corresponds to the vortex density. With the help of the Jacobi Theta function, such a wave function $\Psi$ can be constructed and has been used since the very early study of vortex lattice in type-II superconductor by Abrikosov \cite{Abrikosov} and has been also successfully applied to study two-component BEC \cite{HoMueller}. We shall not repeat the detail of the Jacobi theta function here, but just mention a few key properties for the later use. More details can be found in Ref. \cite{HoMueller}.

The Fourier transformation of $|\Psi|^2$ is
\begin{equation}
|\Psi|^2=\frac{1}{f_{00}}\left[\sum_{mn}f_{mn}e^{i{\bf K}{\bf r}}\right]e^{-{\bf r}^2/\sigma^2}. \label{Psi2}
\end{equation}
Here we denote two reciprocal lattice vectors ${\bf K}_1=2\pi {\bf b}_2\times\hat{{\bf z}}/v_\text{c}=(2\pi/a)(\hat{x}-u\hat{y}/v)$ and ${\bf K}_2=-(2\pi/b_2)\hat{y}/v$, and ${\bf K}=m{\bf K}_1+n{\bf K}_2$, where $m$ and $n$ are both integers. And
\begin{equation}
f_{{\bf K}}=(-1)^{m+n+mn}e^{-v_\text{c}|{\bf K}|^2/(8\pi)}\sqrt{\frac{v_\text{c}}{2}}.
\end{equation}
One can see from here that once $u$, $v$ and $v_\text{c}$ (or equivalently $a$) are given and the lattice structure is fixed, $f_{{\bf K}}$ is also fixed and the entire function form of Eq. \ref{Psi2} is determined. $A_\perp=\pi\sigma^2$ is the size of the cloud in the $xy$ plane, considering a uniform system, and in the limit $\sigma\rightarrow\infty$ (in practices $\sigma^2\gg v_\text{c}$), it is easy to show that $\int d^2{\bf r}|\Psi|^2=1$.  The integration of $\int d^2{\bf r}|\Psi|^4$ is denoted by $I/(2A_\perp)$ and
\begin{equation}
I=\sum\limits_{{\bf K}}\left|\frac{f_{{\bf K}}}{f_{00}}\right|^2. \label{I}
\end{equation}

Now we take the \textit{ansatz} for $\Delta({\bf r})$ and $\varphi({\bf r})$ as 
\begin{align}
&\Delta({\bf r})=\Delta_0\sqrt{A_\perp}\Psi_\text{f},
&\varphi({\bf r})=\sqrt{n_{2d}}\Psi_\text{b},
\end{align}
where $n_{\text{2d}}$ is the two-dimensional density of bosons, which relates to boson density $n_\text{b}$ via $n_\text{b}=n_\text{2d}/A_\perp$. $\Psi_\text{f}$ and $\Psi_\text{b}$ have the same function form as $\Psi$ discussed above, but they generally can have different lattice structure described by two different sets of parameters $\{ u^\text{f}, v^\text{f}, v^\text{f}_\text{c}\}$ and $\{ u^\text{b}, v^\text{b}, v^\text{b}_\text{c}\}$, respectively, and correspondingly, different reciprocal lattice vectors denoted by ${\bf K}^\text{f}$ and ${\bf K}^\text{b}$, and different $f$-functions denoted by $f^\text{f}_{{\bf K}}$ and $f^\text{b}_{{\bf K}}$, respectively. $I_\text{f}$ and $I_\text{b}$ denote Eq. \ref{I} with $f_{{\bf K}}$ being $f^\text{f}_{{\bf K}}$ and $f^\text{b}_{{\bf K}}$, respectively. Finally, the integration $\int d^2{\bf r} |\Psi^\text{f}|^2|\Psi^\text{b}|^2$ is denoted by $I_\text{bf}/(2A_\perp)$, where
\begin{equation}
I_\text{bf}=\sum_{{\bf K}{\bf K^\prime}}\frac{f^\text{f}_{{\bf K}}f^\text{b}_{{\bf K}^\prime}}{f^\text{f}_{00}f^\text{b}_{00}}e^{-i{\bf K}\cdot{\bf r}_0}\delta({\bf K}+{\bf K}^\prime), \label{IBF}
\end{equation}
where $r_0$ is the relative displacement between two lattices. 

\textit{Minimization of the Free-Energy.} Based on the aforementioned properties of $\Psi$, the terms proportional to $|\Psi_\text{f}|^2$ or $|\Psi_\text{b}|^2$ simply contribute a constant to the free-energy, and only the terms proportional to $|\Psi_\text{f}|^4$, $|\Psi_\text{b}|^4$ or $|\Psi_\text{f}|^2|\Psi_\text{b}|^2$ depend on the lattice structure parameters. In $\mathcal{F}_\text{f}$, the term is
\begin{align}
\mathcal{E}_\text{F}\alpha_2\int d^3{\bf r}\left|\frac{\Delta({\bf r})}{\Delta_0}\right|^4=\frac{\mathcal{V}}{2}\mathcal{E}_\text{F}\alpha_2 I_\text{f}
\end{align}
where $\mathcal{V}$ is the total volume of the system. In $\mathcal{F}_\text{b}$, the term is
\begin{equation}
\frac{2\pi\hbar^2 a_\text{bb}}{m_\text{b}}\int d^3{\bf r}|\varphi({\bf r})|^4=\frac{\mathcal{V}}{2}\frac{2\pi\hbar^2 a_\text{bb}}{m_\text{b}}n^2_\text{b}I_\text{b}.
\end{equation}
And finally in $\mathcal{F}_\text{bf}$, the term is
\begin{equation}
\frac{2\pi\hbar^2 a_\text{bf}n_\text{f}\beta_1}{m_\text{bf}}\int d^3{\bf r}\left|\frac{\Delta({\bf r})}{\Delta_0}\right|^2|\varphi({\bf r})|^2=\frac{\mathcal{V}}{2}\frac{2\pi\hbar^2 a_\text{bf}n_\text{f}n_\text{b}\beta_1}{m_\text{bf}}I_\text{bf}.
\end{equation}

Now we introduce two dimensionless parameters 
\begin{align}
& \bar{g}_\text{b}=\frac{2\pi\hbar^2 a_\text{bb}}{m_\text{b}\mathcal{E}_\text{F}}n^2_\text{b};
& \bar{g}_\text{bf}=\frac{2\pi\hbar^2 a_\text{bf}}{m_\text{bf}\mathcal{E}_\text{F}}n_\text{f}n_\text{b}.
\end{align}
Then, to determine the vortex structure, we only need to minimize the following free-energy density 
\begin{equation}
F=\alpha_2 I_\text{f}+\bar{g}_\text{b}I_\text{b}+\bar{g}_\text{bf}\beta_1I_\text{bf}. \label{F}
\end{equation}
This is a key equation for this work. Then, quite straightforward numerical evaluation will yield the lattice structure for different interaction parameters. 

\textit{Results.} First of all, we find that, for equal vortex densities of two superfluids, the interaction between the Bose and the Fermi superfluids, (i.e. the $I_\text{bf}$ term in Eq. \ref{F}), always locks the structure of two sets of vortex lattice to be identical. This is because, when ${\bf K}$ and ${\bf K}^\prime$ match each other, the $\cos({\bf K}\cdot{\bf r}_0)$ term in Eq. \ref{IBF} can always lower the energy by choosing ${\bf r}_0$ properly. Generalizing to other vortex density ratio is also quite straightforward.

\begin{figure}[tp]
\includegraphics[width=3.0 in]
{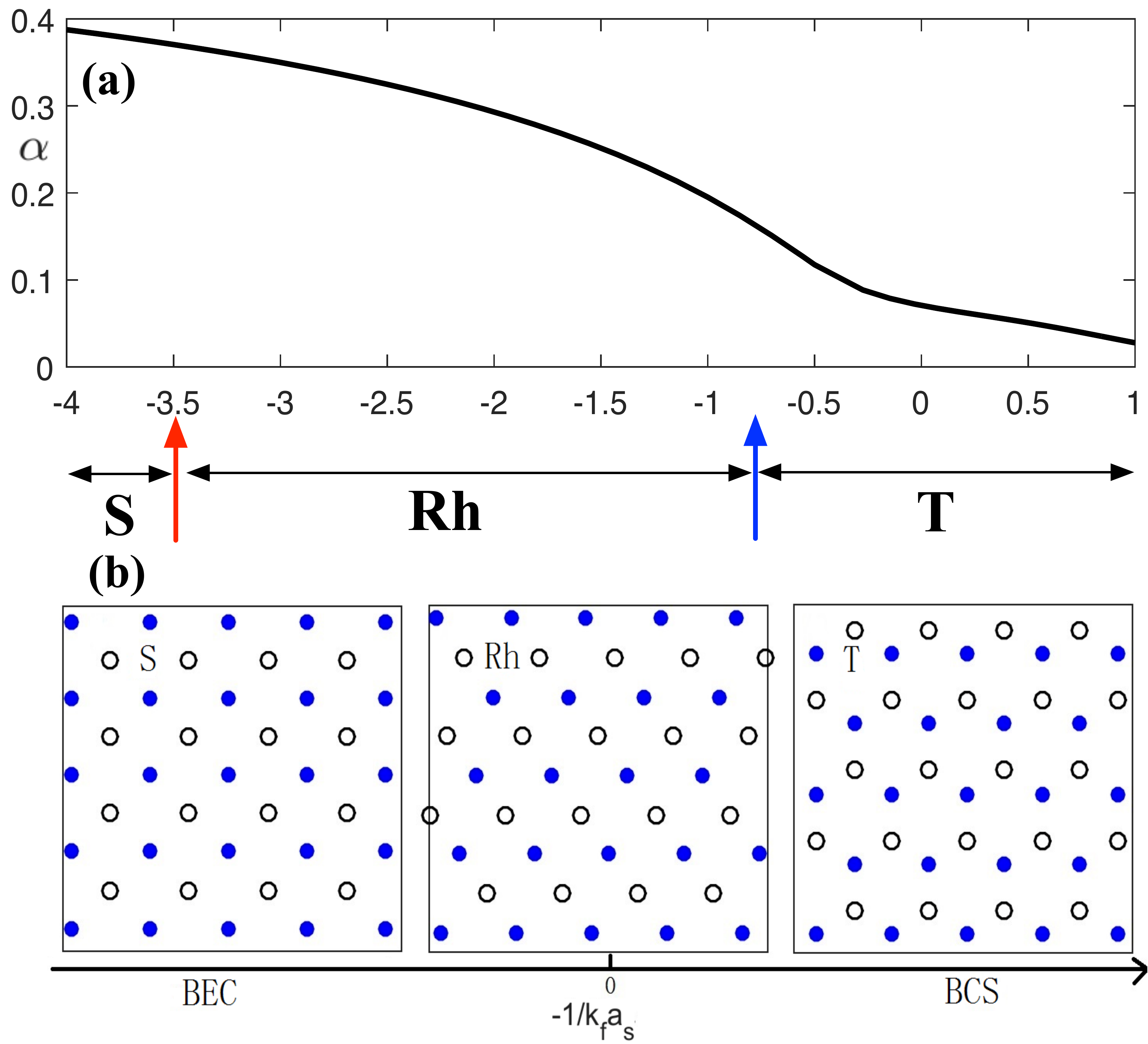}
\caption{Vortex lattice structure in the BEC-BCS crossover of the Fermi superfluid. Here we have taken $\bar{g}_\text{b}=0.14$ and $\bar{g}_\text{bf}=0.08$. $\alpha=\bar{g}_\text{bf}\beta_1/(\alpha_2+\bar{g}_\text{b})$. The arrows label the transition from triangular lattice (``T") to rhombic lattice (``Rh"), and from rhombic lattice (``Rh") to square lattice (``S").  }
\label{Lattice}
\end{figure}

For identical lattice structures, $I_\text{f}=I_\text{b}$, and therefore the structure only depends on the ratio $\alpha=\bar{g}_\text{bf}\beta_1/(\alpha_2+\bar{g}_\text{b})$. When $\alpha\ll 1$, $I_\text{f}$ and $I_\text{b}$ terms are dominative, which favor two sets of triangular lattice (denoted by ``T''); while when $\alpha$ increases, $I_\text{bf}$ will becomes dominating, and it will drive a sequence of structure transitions: first to a rhombic lattice (denoted by ``Rh"), and then to a square lattice (denoted by ``S") and eventually to a rectangular lattice (denoted by ``Re"). As shown in Fig. \ref{Lattice}(a), we find that $\alpha$ increases monotonically from the BCS side to the BEC side, when $\bar{g}_\text{b}$ and $\bar{g}_\text{bf}$ are fixed at typical experimental values. Hence, this exactly manifests the physics mentioned at the introduction, i.e. these two sets of vortex lattices interact stronger at the BEC regime. As shown in Fig. \ref{Lattice}(b), a sequence of structure transitions are driven by tuning $-1/(k_\text{F}a_\text{s})$.

\begin{figure}[tp]
\includegraphics[width=3.2 in]
{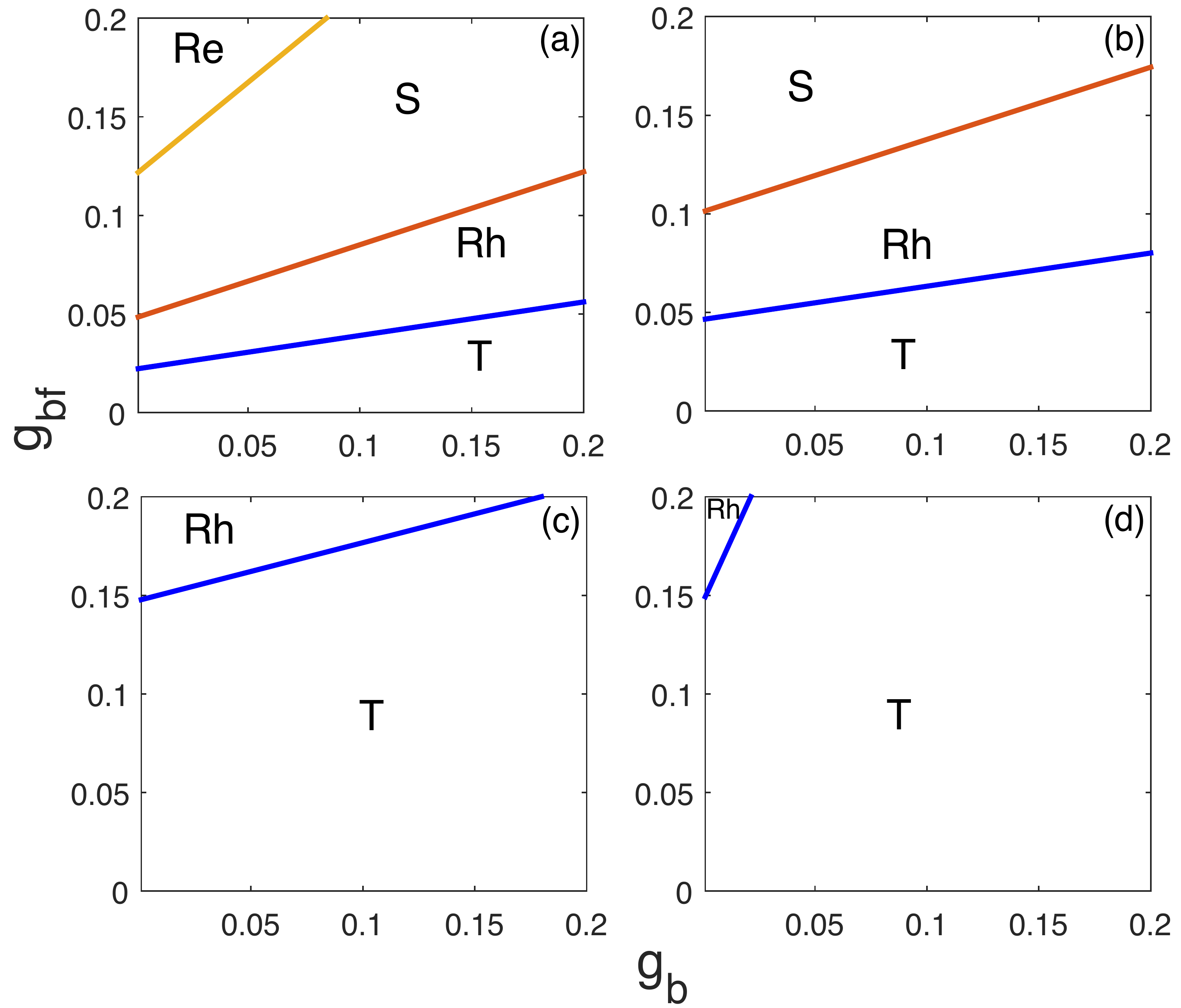}
\caption{The phase diagram for vortex lattice as a function of $\bar{g}_\text{b}$ and $\bar{g}_\text{bf}$ for $1/(k_\text{F}a_\text{s})=2.0$ (a); $=1.0$(b); $=0$(c) and $=-1$(d). ``T", ``Rh", ``S" and ``Re" denote triangular, rhombic, square and rectangular lattices, respectively.  }
\label{diagram}
\end{figure}

Similar physics is also shown alternatively in Fig. \ref{diagram}, where the phase diagrams for vortex lattice is plotted in term of $\bar{g}_\text{b}$ and $\bar{g}_\text{bf}$ for various fixed value of $-1/(k_\text{F}a_\text{s})$. In Fig. \ref{diagram}(a-b), the system is in the BEC side and it is easier for $\bar{g}_\text{bf}$ to drive structure transitions. In Fig. \ref{diagram}(a), a reasonable value of $\bar{g}_\text{bf}$ can even drive the system into rectangular lattice (Re) regime. While at the unitary regime (Fig. \ref{diagram}(c)) and the BCS regime (Fig. \ref{diagram}(d)), it becomes more and more difficult to drive the phase transition and most regime of the phase diagram is occupied by the triangular lattice (T) phase. 

\textit{Final Remarks.} Finally we shall remark that the phenomenon discovered here is a direct manifestation of different characters of the Fermi superfluid at the BCS and the BEC side, and therefore, is unique to this new superfluid mixture. Because various approximations have been implemented in order to obtain the results without involving heavy numerics, our phase diagram is not quantitively accurate but it should be qualitatively correct, due to the robustness of the underlying physics. Given the experimental progresses on this system, our work will stimulate more efforts in investigating this system. 

\textit{Acknowledgment.} This work is supported by NSFC Grant No. 11325418 (HZ), No. 11304357(YZJ), No. 11534014(YZJ), Tsinghua University Initiative Scientific Research Program (HZ), MOST under Grant No. 2016YFA0301600(HZ), the Fundamental Research Funds for the Central Universities(RQ), and the Research Funds of Renmin University of China under Grant No. 15XNLF18 (RQ) and No. 16XNLQ03 (RQ).

\begin{thebibliography}{99}

\bibitem{Abrikosov}
A. A. Abrikosov, Sov. Phys. JETP 5, 1174 (1957)

\bibitem{Dalibard}
K. W. Madison, F. Chevy, W. Wohlleben, and J. Dalibard, Phys. Rev. Lett. {\bf 84}, 806 (2000)

\bibitem{KetterleBEC}
J. R. Abo-Shaeer, C. Raman, J. M. Vogels, W. Ketterle, Science, {\bf 292}, 476 (2001)

\bibitem{review}
For a review, see A. L. Fetter, Rev. Mod. Phys. {\bf 81}, 647 (2009)

\bibitem{synthetic}
Y. J. Lin, R, L. Compton, K. J. Garcia, J. V. Porto, I. B. Spielman, Nature {\bf 462}, 628 (2009)

\bibitem{KetterleFermion}
M.W. Zwierlein, J.R. Abo-Shaeer, A. Schirotzek, C.H. Schunck, W. Ketterle, Nature {\bf 435}, 1047 (2005)


\bibitem{HoMueller}
E. J. Mueller and T. L. Ho, Phys. Rev. Lett. {\bf 88}, 180403 

\bibitem{Ueda}
K. Kasamatsu, M. Tsubota, and M. Ueda Phys. Rev. Lett. {\bf 91}, 150406 (2003)

\bibitem{JILA}
V. Schweikhard, I. Coddington, P. Engels, S. Tung, and and E. Cornell, Phys. Rev. Lett. {\bf 93} 210403 (2004)

\bibitem{Salamon}
I. Ferrier-Barbut, M. Delehaye, S. Laurent, A. T. Grier, M. Pierce, B. S. Rem, F. Chevy and C. Salomon, Science, 345, 1035 (2014)

\bibitem{Salamon2}
M. Delehaye, S. Laurent, I. Ferrier-Barbut, S. Jin, F. Chevy, C. Salomon,  Phys. Rev. Lett. {\bf 115}, 265303 (2015)

\bibitem{Yuao}
X. C. Yao, H. Z. Chen, Y. P. Wu, X. P. Liu, X. Q. Wang, X. Jiang, Y. J. Deng, Y. A. Chen and J. W. Pan, arXiv: 1606.01717

\bibitem{Zheng}
W. Zheng and H. Zhai, Phys. Rev. Lett. {\bf 113}, 265304 (2014)

\bibitem{Zhang}
R. Zhang, W. Zhang, H. Zhai and P. Zhang, Phys. Rev. A {\bf 90}, 063614 (2014)

\bibitem{Cui}
X. Cui, Phys. Rev. A 90, 041603(R) (2014)

\bibitem{Stringari}
T. Ozawa, A. Recati, M. Delehaye, F. Chevy, and S. Stringari, Phys. Rev. A 90, 043608 (2014)

\bibitem{mean-field}
S. Giorgini, L. P. Pitaevskii, S. Stringari, Rev. Mod. Phys. {\bf 80}, 1215 (2008)



\end{thebibliography}
\end{document}